\documentstyle[11pt]{article}
\bibliography{plain}
\title{\bf Optimal Lewenstein-Sanpera Decomposition for some
Biparatite Systems} \vspace{20mm}
\author{
 S. J. Akhtarshenas$^{a,b,c}$
\thanks{E-mail:akhtarshenas@tabrizu.ac.ir}
 , M. A. Jafarizadeh$^{a,b,c}$ \thanks{E-mail:jafarizadeh@tabrizu.ac.ir}
\\
\\
$^a${\small Department of Theoretical Physics and Astrophysics,
Tabriz University, Tabriz 51666, Iran.} \\
$^b${\small Institute for Studies in Theoretical Physics and Mathematics,
 Tehran 19395-1795, Iran.} \\
$^c${\small Research Institute for Fundamental Sciences, Tabriz
51666, Iran.}} \pagebreak

\begin{document}
\maketitle \vspace{15mm}
\newpage
\begin{abstract}
It is shown that for a given bipartite density matrix and by
choosing a suitable separable set (instead of product set) on the
separable-entangled boundary, optimal Lewenstein-Sanpera (L-S)
decomposition can be obtained via optimization for a generic
entangled density matrix. Based on this, We obtain optimal L-S
decomposition for some bipartite systems such as $2\otimes 2$ and
$2\otimes 3$ Bell decomposable states, generic two qubit state in
Wootters basis, iso-concurrence decomposable states, states
obtained from BD states via one parameter and three parameters
local operations and classical communications (LOCC), $d\otimes d$
Werner and isotropic states, and a one parameter $3\otimes 3$
state. We also obtain the optimal decomposition for multi partite
isotropic state.  It is shown that in all $2\otimes 2$ systems
considered here the average concurrence of the decomposition is
equal to the concurrence. We also show that for some $2\otimes 3$
Bell decomposable states the average concurrence of the
decomposition is equal to the lower bound of the concurrence of
state presented recently in [Buchleitner et al, quant-ph/0302144],
so an exact expression for concurrence of these states is
obtained. It is also shown that for $d\otimes d$ isotropic state
where decomposition leads to a separable and an entangled pure
state, the average I-concurrence of the decomposition is equal to
the I-concurrence of the state.


{\bf PACS Index: 03.65.Ud }
\end{abstract}
\pagebreak

\vspace{7cm}

\section{Introduction}\label{secInt}
In the past decade quantum entanglement has been attracted much
attention in connection with theory of quantum information and
computation. This is because of potential resource that
entanglement provides for quantum communication and information
processing \cite{ben1,ben2,ben3}. By definition, a bipartite mixed
state $\rho$ is said to be entangled if it can not be expressed as
$$ \rho=\sum_{i}w_{i}\,\rho_i^{(1)}\otimes\rho_i^{(2)},\qquad
w_i\geq 0, \quad \sum_{i}w_i=1, $$ where $\rho_i^{(1)}$ and
$\rho_i^{(2)}$ denote density matrices of subsystems 1 and 2,
respectively. Otherwise the state is separable.

The central tasks of quantum information theory is to characterize
and quantify entangled states. A first attempt in characterization
of entangled states has been made by Peres and Horodecki et al.
\cite{peres,horo0}. Peres showed that a necessary condition for
separability of a bipartite system is that its partial transpose
be positive. Horodecki et al. have shown that this condition is
sufficient for separability of composite systems only for
dimensions $2\otimes 2$ and $2 \otimes 3$.

Having a well justified measure to quantify entanglement,
particularly for mixed states of a bipartite system, is indeed
worth, and a number of measures have been proposed
\cite{ben3,ved1,ved2,woot}. Among them the entanglement of
formation has more importance, since it intends to quantify the
resources needed to create a given entangled state.

Another interesting description of entanglement is
Lewenstein-Sanpera (L-S) decomposition \cite{LS,karnas}.
Lewenstein and Sanpera have shown that any bipartite density
matrix can be represented optimally as a sum of a separable state
and an entangled state. They have also shown that for two qubit
systems the decomposition reduces to a mixture of a mixed
separable state and an entangled pure state, thus all entanglement
content of the state is concentrated in the pure entangled state.
This leads to an unambiguous measure of entanglement for any two
qubit state as entanglement of pure state multiplied by the weight
of pure part in the decomposition. The strategy of Refs.
\cite{LS,karnas} is based on the fact that for a given set
$V=\{\left|e_\alpha,f_\alpha\right>\}$ of product states belonging
to the range of density matrix $\rho$, one can subtract separable
density matrix $\rho_s^\ast=\sum_\alpha\Lambda_\alpha P_\alpha$
(not necessary normalized) with $\Lambda_\alpha\ge 0$ such that
$\delta \rho=\rho-\rho_s^\ast\ge 0$.

In Ref. \cite{LS}, the best separable approximation (BSA) has been
obtained numerically in case of two qubit Werner state by choosing
a set of several hundred $P_\alpha$-projectors. Some analytical
results is also obtained for special states of two qubit states
\cite{englert}. Further, in \cite{kus} BSA of a two qubit state
has been obtained algebraically. They have also shown that in some
cases the weight of the entangled part in the decomposition is
equal to the concurrence of the state. In Ref. \cite{jaf1} we have
obtained optimal L-S decomposition for a generic two qubit density
matrix by using Wootters basis. It is shown that the average
concurrence of the decomposition is equal to the concurrence of
the state.

In this paper we obtain optimal L-S decomposition for some
bipartite systems. Here we obtain optimal decomposition for a
given density matrix $\rho$ by choosing suitable separable set $S$
in which $\rho_s\in S$. This approach is different from the others
in the sense that optimal decomposition is obtained for a given
separable set $S$ instead of product set $V$. Also this approach
is geometrically intuitive as it will be explained in section 4 by
providing a bunch of  interesting bipartite systems such as,
$2\otimes 2$ and $2\otimes 3$ Bell decomposable states, a generic
two qubit state in Wootters basis, iso-concurrence decomposable
states, states differing from BD states via one parameter and
three parameters local operations and classical communications
(LOCC), $d\otimes d$ Werner and isotropic states, and a one
parameter $3\otimes 3$ state.  We also provide the optimal
decomposition for multi partite isotropic system.  As a byproduct
we show that in all $2\otimes 2$ systems considered here the
average concurrence of the decomposition is equal to the
concurrence. We also show that for some $2\otimes 3$ Bell
decomposable states for which entangled part of the decomposition
is only a pure state the average concurrence of the decomposition
is equal to the lower bound of the concurrence of state presented
recently in Ref. \cite{loz}, consequently  an exact expression for
concurrence of these states is given. In the case of $d\otimes d$
isotropic state we show that the average I-concurrence of the
decomposition is equal to the I-concurrence of the state.

The paper is organized as follows. In section 2 we, briefly,
review concurrence as presented in \cite{woot}. In section 3 we
first review  Lewenstein-Sanpera decomposition for bipartite
density matrix, then a new prescription for finding optimal
decomposition is presented. Some important bipartite examples is
considered in section 4. The paper is ended with a brief
conclusion in section 5.

\section{Concurrence}\label{secConcurrence}
In this section we review concurrence of two qubit mixed states as
introduced in \cite{woot}. The generalized concurrence is also
reviewed, briefly.
\subsection{Wootters's Concurrence}\label{subsecWootcon}
From the various measures proposed to quantify entanglement, the
entanglement of formation has a special position which in fact
intends to quantify the resources needed to create a given
entangled state \cite{ben3}. Wootters in \cite{woot} has shown
that for a two qubit system entanglement of formation of a mixed
state $\rho$ can be defined as
\begin{equation}\label{EoF2}
E_f(\rho)=H\left(\frac{1}{2}+\frac{1}{2}\sqrt{1-C^2}\right),
\end{equation}
where $H(x)=-x\ln{x}-(1-x)\ln{(1-x)}$ is binary entropy and
concurrence $C(\rho)$ is defined by
\begin{equation}\label{concurrence}
C(\rho)=\max\{0,\lambda_1-\lambda_2-\lambda_3-\lambda_4\},
\end{equation}
where the $\lambda_i$ are the non-negative eigenvalues, in
decreasing order, of the Hermitian matrix
$R\equiv\sqrt{\sqrt{\rho}{\tilde \rho}\sqrt{\rho}}$ where the
spin-flipped state ${\tilde \rho}$ is defined by
\begin{equation}\label{rhotilde}
{\tilde \rho}
=(\sigma_y\otimes\sigma_y)\rho^{\ast}(\sigma_y\otimes\sigma_y),
\end{equation}
where $\rho^{\ast}$ is the complex conjugate of $\rho$ in a
standard basis such as $\{\left|00\right>, \left|01\right>,
\left|10\right>, \left|11\right>\}$ and $\sigma_y$ represent Pauli
matrix in local basis $\{\left|0\right>, \left|1\right>\}$.

Consider a generic two qubit density matrix $\rho$ with its
subnormalized orthogonal eigenvectors $\left|v_i\right>$, i.e.
$\rho=\sum_i \left|v_i\right>\left<v_i\right|$. There always exist
a decomposition \cite{woot}
\begin{equation}\label{rhox}
\rho=\sum_i\left|x_i\right>\left<x_i\right|
\end{equation}
where Wootters's basis $\left|x_i\right>$ are defined by
\begin{equation}\label{xvector}
\left|x_i\right>=\sum_{j}^{4}U_{ij}^{\ast}\left|v_i\right>, \qquad
\mbox{for}\quad i=1,2,3,4,
\end{equation}
such that
\begin{equation}\label{xortho}
\left<x_i\mid \tilde{x}_j\right>=(U\tau
U^T)_{ij}=\lambda_i\delta_{ij},
\end{equation}
where $\tau_{ij}=\left<v_i\mid \tilde{v}_j\right>$ is a symmetric
but not necessarily Hermitian matrix. The states
$\left|x_i^{\prime}\right>$, which  are going to be used in our
notation, is defined as
\begin{equation}
\left|x_i^{\prime}\right>=\frac{\left|x_i\right>}{\sqrt{\lambda_i}},\qquad
\rm{for}\quad i=1,2,3,4.
\end{equation}

\subsection{I-concurrence}\label{subsecIcon}
Several attempts to generalize the notion of concurrence for
arbitrary bipartite quantum system have been made already
\cite{uhlmann,bad,rungta1}. Among them the so-called I-concurrence
\cite{rungta1} is defined in terms of universal-inverter
superoperator which is a natural generalization to higher
dimensions of two qubit spin flip. I-concurrence of a joint pure
state $\left|\psi\right>$ of a $d_A \otimes d_B$ system is defined
by Rungta et al. \cite{rungta1}
\begin{equation}\label{Icon}
C(\left|\psi\right>)=\sqrt{2(1-\rm{tr}(\rho_A^2))}
=\sqrt{2(1-\rm{tr}(\rho_B^2))},
\end{equation}
where $\rho_A=\rm{tr}_B(\left|\psi\right>\left<\psi\right|)$ and
$\rho_B$ is defined similarly.

\section{Lewenstein-Sanpera decomposition}\label{secLS}
According to Lewenstein-Sanpera decomposition \cite{LS}, any
bipartite  density matrix $\rho$ can be written as
\begin{equation}\label{LSD}
\rho=\lambda\rho_{s}+(1-\lambda)\rho_e, \quad\quad
\lambda\in[0,1],
\end{equation}
where $\rho_{s}$ is a separable density matrix and $\rho_e$ is an
entangled state. The Lewenstein-Sanpera (L-S) decomposition of a
given density matrix $\rho$ is not unique and, in general, there
is a continuum set of L-S decomposition to choose from. However,
Lewenstein and Sanpera in \cite{LS,karnas} have shown that the
optimal decomposition is unique for which $\lambda$ is maximal.
Furthermore they have demonstrated that in the case of two qubit
systems $\rho_e$ reduces to a single pure state.

The idea of Refs. \cite{LS,karnas} is based on the method of
subtracting projections on product vectors from a given state,
that is, for a given density matrix $\rho$ and any set
$V=\{\left|e_\alpha,f_\alpha\right>\}$ of product states belonging
to the range of $\rho$, one can subtract separable density matrix
$\rho_s^\ast=\sum_\alpha\Lambda_\alpha P_\alpha$ (not necessary
normalized) with all $\Lambda_\alpha\ge 0$ such that $\delta
\rho=\rho-\rho_s^\ast\ge 0$. Separable state $\rho_s^\ast$
provides the optimal separable approximation (OSA) in the sense
that trace $Tr(\rho_s^\ast)\le 1$ is maximal and entangled part
$\rho_e$ is called edge state, a state with no product vectors in
the range \cite{eckert}. Lewenstein and Sanpera provide the
conditions that trace $Tr(\rho_s^\ast)$ is maximal.

In this paper we will deal with L-S decomposition from different
point of view. Our approach is based on the fact that the set of
separable density matrices is convex and compact
\cite{horo1,pitt}. This follows from the fact that any separable
density matrix $\rho_s\in {\cal S}$ can be written as a finite
convex combination of pure product states. The set of all
Hermitian operators acting on the Hilbert space
${\cal{H}}_1\otimes{\cal{H}}_2$ constitute a Hilbert space (called
Hilbert-Shcmidt space) with a real inner product $<A,B>=Tr(A^\dag
B)$. The set of density matrices $\rho$ are defined as Hermitian
positive semi-definite and the trace one matrices  form subset
${\cal D}$ of H-S space which is compact and convex \cite{pitt}.
Let ${\cal P}$ denotes the set of all pure product states. ${\cal
P}$ is tensor product of two spheres which are compact in the
finite dimensional case. So ${\cal P}$ is also compact
\cite{horo1}. The set of all finite convex combinations of product
states ${\cal P}$ is defined as the convex hull of ${\cal P}$,
i.e. ${\cal S}=\mbox{conv}\,{\cal P}$, and convex hull of a
compact set ${\cal P}$ is also compact, so the set of separable
density matrices is compact \cite{horo1}.

Based on the above fact we obtain optimal L-S decomposition for
some bipartite systems. For a given density matrix $\rho$, we
choose a suitable separable set $S\subset {\cal S}$ on the
separable-entangled boundary, and express $\rho$ as a convex
combination of separable state $\rho_s\in S$ and an arbitrary
entangled state $\rho_e$, i.e.
$\rho=\lambda\rho_s+(1-\lambda)\rho_e$. Then we evaluate $\lambda$
and provide the conditions that $\lambda$ is maximal under the
restrictions that $\rho_s$ is in the separable set $S$ and
maintaining the positivity of the difference $\rho-\lambda\rho_s$,
i.e. $\rho_e$ remains nonnegative. To this aim we allow $\rho_s$
to move on the surface defined by $S$, and simultaneously search
for the $\rho_e$ with corresponding maximal $\lambda$. This
restricts $\rho_e$ to some entangled states and gives $\rho_s$ as
a function of $\rho$ and restricted $\rho_e$. The only matter that
should be noticed in choosing the set $S$ for which $\rho_s\in S$,
is that all states on the line segment connecting $\rho_s$ and
$\rho$, i.e. $\rho_\epsilon=\epsilon\rho_s+(1-\epsilon)\rho$ for
$0\le \epsilon\le 1$, must be entangled. This guarantees that thus
obtained decomposition is indeed maximal. In all examples
considered in this paper we will see that the rank of $\rho_e$ is
less than rank of $\rho$. This means that $\rho_e$ is an edge
state with no product vectors in its range as pointed out in Ref.
\cite{eckert}. Moreover in the case of two qubit system it is
shown that $\rho_e$ reduces to pure entangled state as we expect
from the results of Refs. \cite{LS,karnas}. For these systems
$\rho_s$ is defined as a function of $\rho$ and concurrence of
entangled pure state. To make our consideration more clear, we
provide some examples in the next section.

\section{Some important examples}\label{secExam}
In this section we obtain optimal decomposition for some
categories of states, namely, $2\otimes 2$ Bell decomposable (BD)
states, a generic two qubit state in Wootters basis,
iso-concurrence decomposable states, some $2 \otimes 2$ states
obtaining from BD states via one parameter and three parameters
LOCC operations, $2\otimes 3$ Bell decomposable states, $d\otimes
d$ Werner and isotropic states, a one parameter $3\otimes 3$ state
and finally multi partite isotropic state.

\subsection{$2 \otimes 2$ Bell decomposable states}\label{subsecBDS}
We begin by considering the $2\otimes 2$ Bell
decomposable (BD) states. A BD state acting on $H^4\cong
H^2\otimes H^2$ Hilbert space is defined by
\begin{equation}
\rho=\sum_{i=1}^{4}p_{i}\left|\psi_i\right>\left<\psi_i\right|,\quad\quad
0\leq p_i\leq 1,\quad \sum_{i=1}^{4}p_i=1,
 \label{BDS1}
\end{equation}
where $\left|\psi_i\right>$ are Bell states given by
\begin{eqnarray}
\label{BS12} \left|\psi_1\right>=\left|\phi^{+}\right>
=\frac{1}{\sqrt{2}}(\left|00\right>+\left| 11\right>),\qquad
\left|\psi_2\right>=\left|\phi^{-}\right>
=\frac{1}{\sqrt{2}}(\left|00\right>-\left|
11\right>), \\
\label{BS34}\left|\psi_3\right>=\left|\psi^{+}\right>
=\frac{1}{\sqrt{2}}(\left|01\right>+\left| 10\right>),\qquad
\left|\psi_4\right>=\left|\psi^{-}\right>
=\frac{1}{\sqrt{2}}(\left|01\right>-\left| 10\right>).
\end{eqnarray}
A BD state is separable iff $p_i\le \frac{1}{2}$ for all
$i=1,2,3,4$ \cite{horo}. In the following we consider the case
that $\rho$ is entangled for which $p_1 > \frac{1}{2}$. To obtain
optimal L-S decomposition we choose
$\rho_s=\sum_{i=1}^{4}p_{i}^\prime\left|\psi_i\right>\left<\psi_i\right|$
with $p_1^\prime=\frac{1}{2}$ as boundary separable state and
$\rho_e=\sum_{i=1}^{4}p_{i}^{\prime\prime}\left|\psi_i\right>\left<\psi_i\right|$.
Inserting these equations into the decomposition given in Eq.
(\ref{LSD}) we get
\begin{equation}\label{BDLSD1}
p_i=\lambda p_i^{\prime}+(1-\lambda)p_i^{\prime\prime}\qquad
\rm{for}\quad i=1,2,3,4.
\end{equation}
From Eq. (\ref{BDLSD1}) we get
$\lambda=\frac{C^{\prime\prime}-C}{C^{\prime\prime}}$ and
$\frac{\rm{d}\lambda}{\rm{d} C^{\prime\prime}}
=\frac{C}{{C^{\prime\prime}}^2}\ge 0$ where  $C=2p_1-1$ and
$C^{\prime\prime} =2p_1^{\prime\prime}-1$ are concurrence of
$\rho$ and $\rho_e$, respectively. This means that in order to
obtain optimal decomposition, i.e. having maximal $\lambda$, we
should require that $C^{\prime\prime}$ takes its maximal value,
where this happens as long as
$p_2^{\prime\prime}=p_3^{\prime\prime}=p_4^{\prime\prime}=0$, i.e.
$\rho_e$ is pure entangled state. Considering the above arguments
we get for $\lambda$, $\rho_s$ and $\rho_e$ the following results
\begin{equation}\label{lambdaBD}
\begin{array}{c}
\lambda=1-C, \qquad \rho_e=\left|\psi_1\right>\left<\psi_1\right|,
\\
p_1^\prime=\frac{1}{2},\qquad
p_j^{\prime}=\frac{p_j}{\lambda}\quad \mbox{for}\,\,j=2,3,4.
\end{array}
\end{equation}
Equation (\ref{lambdaBD}) simply shows that average concurrence of
the decomposition is equal to the concurrence of state, i.e.
$(1-\lambda)C(\left|\psi\right>)=C$.

\subsection{ A generic two qubit state in Wootters's basis}
\label{subsecGwoot}
In this subsection we obtain optimal L-S decomposition for a
generic two qubit density matrix by using Wootters basis. In Ref.
\cite{jaf1} we have shown that a generic two qubit density matrix
$\rho=\sum_{i}\lambda_i\left|x_i^{\prime}\right>\left<x_i^{\prime}\right|$
with corresponding set of positive numbers $\lambda_i$ and
Wootters's basis $\left|x_i^{\prime}\right>$ can be obtained from
a Bell decomposable state with the same set of positive numbers
$\lambda_i$ but with different Wootters's basis via $SO(4,c)$
transformation. It is also shown that local unitary
transformations correspond to $SO(4,r)$ transformations, hence,
$\rho$ can be represented as coset space $SO(4,c)/SO(4,r)$
together with positive numbers $\lambda_i$.

Now in order to obtain optimal L-S decomposition we choose
$\rho_s=\sum_{i}\lambda_i^{\prime}\left|x_i^{\prime}\right>\left<x_i^{\prime}\right|$
with
$\lambda_1^{\prime}-\lambda_2^{\prime}-\lambda_3^{\prime}-\lambda_4^{\prime}=0$
as boundary separable state and
$\rho_e=\sum_{i}\lambda_i^{\prime\prime}
\left|x_i^{\prime}\right>\left<x_i^{\prime}\right|$. Inserting
these equations into the decomposition given in Eq. (\ref{LSD}) we
get
\begin{equation}\label{wootLSD1}
\lambda_i=\lambda
\lambda_i^{\prime}+(1-\lambda)\lambda_i^{\prime\prime}\qquad
\rm{for}\quad i=1,2,3,4.
\end{equation}
From Eq. (\ref{wootLSD1}) we get
$\lambda=\frac{C^{\prime\prime}-C}{C^{\prime\prime}}$ and
$\frac{\rm{d}\lambda}{\rm{d} C^{\prime\prime}}
=\frac{C}{{C^{\prime\prime}}^2}\ge 0$ where $C=\lambda_1-\lambda_2
-\lambda_3-\lambda_4$ and
$C^{\prime\prime}=\lambda_1^{\prime\prime}-\lambda_2^{\prime\prime}
-\lambda_3^{\prime\prime}-\lambda_4^{\prime\prime}$ are
concurrence of $\rho$ and $\rho_e$, respectively. This means that
in order to obtain optimal decomposition, i.e. having maximal
$\lambda$, we should require that $C^{\prime\prime}$ takes its
maximal value, which happens as long as
$\lambda_2^{\prime\prime}=\lambda_3^{\prime\prime}=\lambda_4^{\prime\prime}=0$,
i.e. $\rho_e$ is pure entangled state with concurrence
$\lambda_1^{\prime\prime}$. Considering the above arguments we get
for $\lambda$, $\rho_s$ and $\rho_e$ the following results
\begin{equation}\label{lambdawoot}
\begin{array}{c}
\lambda=1-\frac{C}{\lambda_1^{\prime\prime}}, \qquad
\rho_e=\lambda_1^{\prime\prime}
\left|x_1^{\prime}\right>\left<x_1^{\prime}\right|,
\\
\lambda_1^{\prime}=\frac{\lambda_2+\lambda_3+\lambda_4}
{\lambda},\qquad \lambda_j^{\prime}=\frac{\lambda_j}{\lambda}\quad
\mbox{for}\,\,j=2,3,4.
\end{array}
\end{equation}
Equation (\ref{lambdawoot}) simply shows that average concurrence
of the decomposition is equal to the concurrence of state, i.e.
$(1-\lambda)C(\left|\psi\right>)=C$. The decomposition
(\ref{lambdawoot}) is in agrement with results obtained in Ref.
\cite{jaf1}

\subsection{Iso-concurrence decomposable states}\label{subsecICD}
In this section we define iso-concurrence decomposable (ICD)
states, then we give their separability condition and evaluate
optimal decompsition. The iso-concurrence states are defined by
\begin{eqnarray} \label{ICS12}
\left|\phi_1\right>=\cos{\theta}\left|00\right>
+\sin{\theta}\left|11\right>),\qquad
\left|\phi_2\right>=\sin{\theta}\left|00\right>
-\cos{\theta}\left|11\right>), \\
\label{ICS34} \left|\phi_3\right>=\cos{\theta}\left|01\right>
+\sin{\theta}\left|10\right>),\qquad
\left|\phi_4\right>=\sin{\theta}\left|01\right>
-\cos{\theta}\left|10\right>).
\end{eqnarray}
It is quite easy to see that the above states are orthogonal thus
span the Hilbert space of $2\otimes 2$ systems. Also by choosing
$\theta=\frac{\pi}{4}$  the above states reduce to Bell states.
Now we can define ICD states as
\begin{equation} \label{ICDS}
\rho=\sum_{i=1}^{4}p_{i}\left|\phi_i\right>\left<\phi_i\right|,\quad\quad
0\leq p_i\leq 1,\quad \sum_{i=1}^{4}p_i=1.
\end{equation}
 These states form a four simplex (tetrahedral)  with its
vertices defined by $p_1=1$, $p_2=1$, $p_3=1$ and $p_4=1$,
respectively.

Peres-Horodeckis criterion \cite{peres,horo0} for separability
implies that the state given in Eq. (\ref{ICDS}) is separable if
and only if the following inequalities are satisfied
\begin{eqnarray}
\label{ppt1}
(p_1-p_2)\leq \sqrt{4p_3p_4+(p_3-p_4)^2\sin^2{2\theta}}, \\
\label{ppt2}
(p_2-p_1)\leq \sqrt{4p_3p_4+(p_3-p_4)^2\sin^2{2\theta}}, \\
\label{ppt3}
(p_3-p_4)\leq \sqrt{4p_1p_2+(p_1-p_2)^2\sin^2{2\theta}}, \\
\label{ppt4} (p_4-p_3)\leq
\sqrt{4p_1p_2+(p_1-p_2)^2\sin^2{2\theta}}.
\end{eqnarray}
Inequalities (\ref{ppt1}) to (\ref{ppt4}) divide tetrahedral of
density matrices to five regions.  Central regions, defined by the
above inequalities, form a deformed octahedral and are separable
states. In four other regions one of the above inequality will not
hold, therefor they represent entangled states. Bellow we consider
entangled states corresponding to violation of inequality
(\ref{ppt1}) i.e. the states which satisfy the following
inequality
\begin{equation}
\label{ICDE1}
(p_1-p_2)>\sqrt{4p_3p_4+(p_3-p_4)^2\sin^2{2\theta}}.
\end{equation}
All other ICD states can be obtain via local unitary
transformations. Now we will obtain  concurrence  of ICD states.
Following the  Wootters protocol given in subsection
\ref{subsecWootcon} we get for the state $\rho$ given in Eq.
(\ref{ICDS})
\begin{equation}\label{tau}
\tau=\left(\begin{array}{cccc}
-p_1\sin{2\theta} & \sqrt{p_1 p_2}\cos{2\theta} & 0 & 0 \\
\sqrt{p_1 p_2} \cos{2\theta} & p_2\sin{2\theta} & 0 & 0 \\
0 & 0 & p_3\sin{2\theta} & -\sqrt{p_3 p_4}\cos{2\theta} \\
0 & 0 & -\sqrt{p_3 p_4} \cos{2\theta} & -p_4\sin{2\theta}
\end{array}\right).
\end{equation}
Now it is easy to evaluate $\lambda_i$ which yields
\begin{equation}\label{lambda1234}
\begin{array}{c}
\lambda_{1,2}=\frac{1}{2}\left(\pm(p_1-p_2)\sin{2\theta}
+\sqrt{4p_1p_2+(p_1-p_2)^2\sin^2{2\theta}}\right), \\
\lambda_{3,4}=\frac{1}{2}\left(\pm(p_3-p_4)\sin{2\theta}
+\sqrt{4p_3p_4+(p_3-p_4)^2\sin^2{2\theta}}\right).
\end{array}
\end{equation}
Thus one can evaluate the concurrence of ICD states as
\begin{equation}\label{CICDS}
C=(p_1-p_2)\sin{2\theta}-\sqrt{4p_3p_4+(p_3-p_4)^2\sin^2{2\theta}}.
\end{equation}
It is worth to note that thus obtained concurrence is equal to the
amount of violation of inequality (\ref{ICDE1}). Note that the
concurrence of an ICD state can be written as
\begin{equation}\label{Acon}
A_{11}-A_{22}-\sqrt{(A_{33}+A_{44})^2-4A_{34}^2},
\end{equation}
where $A_{ij}$ denote matrix representation of ICD state in Bell
basis, that is
\begin{equation}\label{A11-22}
A_{11}=\frac{1}{2}(p_1+p_2+(p_1-p_2)\sin{2\theta}),\qquad
A_{22}=\frac{1}{2}(p_1+p_2-(p_1-p_2)\sin{2\theta}),
\end{equation}
\begin{equation}\label{A33-44}
A_{33}=\frac{1}{2}(p_3+p_4+(p_3-p_4)\sin{2\theta}),\qquad
A_{44}=\frac{1}{2}(p_3+p_4-(p_3-p_4)\sin{2\theta}),
\end{equation}
\begin{equation}\label{A12-34}
A_{12}=\frac{1}{2}(p_1-p_2)\cos{2\theta},\qquad
A_{34}=\frac{1}{2}(p_3-p_4)\cos{2\theta}.
\end{equation}
Now in order to obtain optimal L-S decomposition we parameterize
$\rho_s$ like ICD state with matrix elements $A_{ij}^{\prime}$ (in
Bell basis) which are defined like $A_{ij}$ except for $p_i$ and
$\theta$ which are replaced with $p_i^{\prime}$ and
$\theta^{\prime}$, respectively. We also choose $\rho_e$ similar
to $\rho$ with matrix elements $A_{ij}^{\prime\prime}$
parameterized with $p_i^{\prime\prime}$ and
$\theta^{\prime\prime}$. For simplicity the rank  of $\rho_e$ is
considered to be two, namely
$p_3^{\prime\prime}=p_4^{\prime\prime}=0$. Using these
consideration together with  Eq. (\ref{LSD}) we get
\begin{equation}\label{Aij}
A_{ij}=\lambda A_{ij}^{\prime}+(1-\lambda)A_{ij}^{\prime\prime},
\end{equation}
Taking into account the fact that $\rho_s$ is boundary separable
state with zero concurrence and using Eq. (\ref{Acon}), we get
$\lambda=\frac{C^{\prime\prime}-C}{C^{\prime\prime}}$ and
$\frac{\rm{d}\lambda}{\rm{d}C^{\prime\prime}}$, where $C$ and
$C^{\prime\prime}$ are concurrence of $\rho$ and $\rho_e$,
respectively. Obviously we observe that $\lambda$ becomes maximal
 when $\rho_e$ is a pure entangled state. Considering this fact and
setting $p_2^{\prime\prime}=0$ we arrive at
\begin{equation}\label{ICDeq11}
p_1+p_2+(p_1-p_2)\sin{2\theta}
=\lambda\left(p_1^{\prime}+p_2^{\prime}
+(p_1^{\prime}-p_2^{\prime})\sin{2\theta^{\prime}}\right)
+(1-\lambda)(1+\sin{2\theta^{\prime\prime}}),
\end{equation}
\begin{equation}\label{ICDeq22}
p_1+p_2-(p_1-p_2)\sin{2\theta}
=\lambda\left(p_1^{\prime}+p_2^{\prime}
-(p_1^{\prime}-p_2^{\prime})\sin{2\theta^{\prime}}\right)
+(1-\lambda)(1-\sin{2\theta^{\prime\prime}}),
\end{equation}
\begin{equation}\label{ICDeq33}
p_3+p_4+(p_3-p_4)\sin{2\theta}
=\lambda\left(p_3^{\prime}+p_4^{\prime}
+(p_3^{\prime}-p_4^{\prime})\sin{2\theta^{\prime}}\right),
\end{equation}
\begin{equation}\label{ICDeq44}
p_3+p_4-(p_3-p_4)\sin{2\theta}
=\lambda\left(p_3^{\prime}+p_4^{\prime}
-(p_3^{\prime}-p_4^{\prime})\sin{2\theta^{\prime}}\right),
\end{equation}
\begin{equation}\label{ICDeq12}
(p_1-p_2)\cos{2\theta}
=\lambda(p_1^{\prime}-p_2^{\prime})\cos{2\theta^{\prime}}
+(1-\lambda)\cos{2\theta^{\prime\prime}},
\end{equation}
\begin{equation}\label{ICDeq34}
(p_3-p_4)\cos{2\theta}
=\lambda(p_3^{\prime}-p_4^{\prime})\cos{2\theta^{\prime}}.
\end{equation}
In order to solve above equations we consider two cases
separability.

{\bf i) case 1:}

First let us consider the case that $\theta,\theta^{\prime}\ne
\frac{\pi}{4}$. In this case Eqs. (\ref{ICDeq11}) to
(\ref{ICDeq34}) yield to
\begin{equation}\label{ICDcase1}
\begin{array}{c}
\theta=\theta^{\prime}=\theta^{\prime\prime}, \\
\lambda=1-(p_1-p_2)\sin{2\theta}+\sqrt{(p_3+p_4)(p_5+p_6)},  \\
p_1^{\prime}=\frac{p_1-(1-\lambda)}{\lambda},\qquad
p_j^{\prime}=\frac{p_j}{\lambda}, \quad \mbox{for}\,\,j=2,3,4.
\end{array}
\end{equation}
This case corresponds to results of Ref. \cite{jaf2}.

{\bf ii) case 2:}

Now let us consider the case that $\theta=\frac{\pi}{4}$, i.e.
$\rho$ is Bell decomposable state. The only nontrivial solution of
Eq. (\ref{ICDeq34}) is $p_3^{\prime}=p_4^{\prime}$. Equations
(\ref{ICDeq33}) and (\ref{ICDeq44}) show that this restricts the
density matrix to $p_3=p_4$. Combining all, we arrive at the
following $\rho_s$ for decomposition
\begin{equation}
\tan{2\theta^{\prime}}=\frac{p_1+p_2-1}{C}\tan{2\theta^{\prime\prime}},
\qquad \lambda=\frac{C}{\sin{2\theta^{\prime\prime}}},
\end{equation}
\begin{equation}
p_{1,2}^{\prime}=\frac{1}{2\lambda}
\left(p_1+p_2-\frac{C}{\sin{2\theta^{\prime\prime}}}
\pm\frac{1-p_1-p_2}{\sin{2\theta^{\prime}}}\right)
\end{equation}
\begin{equation}
p_3^{\prime}=p_4^{\prime}=\frac{p_3}{\lambda}.
\end{equation}
where $C=2p_1-1$ is concurrence of $\rho$. The separability of
density matrix $\rho_s$ implies that  $p_i^{\prime}\ge 0$ for all
$i$ (recall that the separability condition has been already
imposed over $\rho_s$ by putting its concurrence equal to zero).
 So $p_i^{\prime}$ should satisfy the following condition
\begin{equation}
\sin{2\theta^{\prime\prime}}\ge \frac{(p_1+p_2)C}{p_1 C+p_2}.
\end{equation}
This condition also guarantees positivity of $\lambda$. It is
worth to emphasis that this case involves the result of Ref.
\cite{shi} as a special case. There authors have obtained the
optimal decomposition for a special kind of BD states, namely a
specific Werner state with $p_1=\frac{5}{8}$ (of course in their
treatment they take singlet state $\left|\psi_4\right>$ as
dominant pure state in Werner state, i.e. $p_4=\frac{5}{8}$).

\subsection{One parameter LOCC operations}\label{subsecOnepara}
A generic two qubit density matrix $\rho$ can be represented in
Bell basis as $\rho=Y\Lambda Y^\dag$ where $Y\in SO(4,c)/SO(4,r)$
and $\Lambda=\mbox{diag}(\lambda_1,\lambda_2,\lambda_3,\lambda_4)$
\cite{jaf1}. Here we consider the case that $Y$ is a one parameter
matrix as
\begin{equation}
Y=\left(\begin{array}{cccc}
\cosh{\theta} & i\sinh{\theta} & 0 & 0 \\
-i\sinh{\theta} & \cosh{\theta} &0 & 0 \\
0 & 0 & 1 & 0 \\
0 & 0 & 0 & 1
\end{array}\right),
\end{equation}
thus
\begin{equation}
\rho=Y\Lambda Y^\dag=\left(
\begin{array}{cccc}
\lambda_1\cosh^2{\theta}+ \lambda_2\sinh^2{\theta} &
i(\lambda_1+\lambda_2)\sinh{\theta}\cosh{\theta}
& 0 & 0 \\
-i(\lambda_1+\lambda_2)\sinh{\theta}\cosh{\theta} &
\lambda_1\sinh^2{\theta}+
\lambda_2\cosh^2{\theta} & 0 & 0 \\
0 & 0 & \lambda_3 & 0 \\
0 & 0 & 0 & \lambda_4
\end{array}\right).
\end{equation}
Obviously, normalization condition leads to
$(\lambda_1+\lambda_2)\cosh{2\theta}+\lambda_3+\lambda_4=1$. We
choose $\rho_s$ in the same form as $\rho$, i.e.
$\rho_s=Y^{\prime}\Lambda^{\prime}{Y^{\prime}}^\dag$ where
$\Lambda^{\prime}=
\mbox{diag}(\lambda_1^{\prime},\lambda_2^{\prime},
\lambda_3^{\prime},\lambda_4^{\prime})$ and $Y^{\prime}$ is
defined as $Y$ but here $\theta$ is replaced with
$\theta^{\prime}$. Now in order to obtain optimal L-S
decomposition we have to get a generic density matrix for
$\rho_e$. After doing so, it can be easily seen that Eq.
(\ref{LSD}) requires that $\rho_e$ has also the same form as
$\rho$ and $\rho_s$, i.e.
$\rho_e=Y^{\prime\prime}\Lambda^{\prime\prime}{Y^{\prime\prime}}^\dag$
where $\Lambda^{\prime\prime}=
\mbox{diag}(\lambda_1^{\prime\prime},\lambda_2^{\prime\prime},
\lambda_3^{\prime\prime},\lambda_4^{\prime\prime})$ and
$Y^{\prime\prime}$ is defined as $Y$ but with
$\theta^{\prime\prime}$ instead of $\theta$. Inserting the above
equations in Eq. (\ref{LSD}) we get
\begin{equation}\label{oneLSD1}
Y\Lambda Y^\dag=\lambda (Y^{\prime}\Lambda^{\prime}
{Y^{\prime}}^\dag)+(1-\lambda)
(Y^{\prime\prime}\Lambda^{\prime\prime} {Y^{\prime\prime}}^\dag).
\end{equation}
Now  multiplying Eq. (\ref{oneLSD1}) by ${Y^{\prime\prime}}^T$ and
${Y^{\prime\prime}}^\ast$, respectively from left and right and
using the orthogonality of $Y^{\prime\prime}$ we get
\begin{equation}\label{oneLSD2}
({Y^{\prime\prime}}^T Y)\Lambda
(Y^\dag{Y^{\prime\prime}}^\ast)=\lambda ({Y^{\prime\prime}}^T
Y^{\prime})\Lambda^{\prime}
({Y^{\prime}}^\dag{Y^{\prime\prime}}^\ast)+(1-\lambda)\Lambda^{\prime\prime},
\end{equation}
where it can be written as
\begin{equation}\label{oneLSD11}
\left(\lambda_1\cosh^2{(\theta-\theta^{\prime\prime})}
+\lambda_2\sinh^2{(\theta-\theta^{\prime\prime})}\right)=\lambda
\left(\lambda_1^{\prime}\cosh^2{(\theta^{\prime}-\theta^{\prime\prime})}+
\lambda_2^{\prime}\sinh^2{(\theta^{\prime}-\theta^{\prime\prime})}\right)+
(1-\lambda)\lambda_1^{\prime\prime},
\end{equation}
\begin{equation}\label{oneLSD22}
\left(\lambda_1\sinh^2{(\theta-\theta^{\prime\prime})}
+\lambda_2\cosh^2{(\theta-\theta^{\prime\prime})}\right)=\lambda
\left(\lambda_1^{\prime}\sinh^2{(\theta^{\prime}-\theta^{\prime\prime})}+
\lambda_2^{\prime}\cosh^2{(\theta^{\prime}-\theta^{\prime\prime})}\right)+
(1-\lambda)\lambda_2^{\prime\prime},
\end{equation}
\begin{equation}\label{oneLSD33}
\lambda_3=\lambda\lambda_3^{\prime\prime}+(1-\lambda)\lambda_3^{\prime\prime},
\end{equation}
\begin{equation}\label{oneLSD44}
\lambda_4=\lambda\lambda_4^{\prime\prime}+(1-\lambda)\lambda_4^{\prime\prime},
\end{equation}
\begin{equation}\label{oneLSD12}
(\lambda_1+\lambda_2)\sinh{2(\theta-\theta^{\prime\prime})}
+\lambda(\lambda_1^{\prime}+\lambda_2^{\prime})
\sinh{2(\theta^{\prime}-\theta^{\prime\prime})}=0
\end{equation}
Subtracting Eqs. (\ref{oneLSD22}), (\ref{oneLSD33}) and
(\ref{oneLSD44})  from Eq. (\ref{oneLSD11}) and using the fact
that $\rho_s$ is boundary separable state, hence having zero
concurrence, i.e.
$\lambda_1^{\prime}-\lambda_2^{\prime}-\lambda_3^{\prime}-\lambda_4^{\prime}=0$,
we get $\lambda=\frac{C^{\prime\prime}-C}{C^{\prime\prime}}, \quad
\frac{\rm{d}\lambda}{\rm{d}C^{\prime\prime}}=\frac{C}{{C^{\prime\prime}}^2}\ge
0$ where $C=\lambda_1-\lambda_2-\lambda_3-\lambda_4$ and
$C^{\prime\prime}=\lambda_1^{\prime\prime}-\lambda_2^{\prime\prime}
-\lambda_3^{\prime\prime}-\lambda_4^{\prime\prime}$, are
concurrence of $\rho$ and $\rho_e$, respectively. This shows that
maximal $\lambda$ is achieved when
$\lambda_2^{\prime\prime}=\lambda_3^{\prime\prime}=\lambda_4^{\prime\prime}=0$,
i.e. $\rho_e$ is pure entangled state with concurrence
$\lambda_1^{\prime\prime}$. Implying the above results we can
solve Eqs. (\ref{oneLSD11}) to (\ref{oneLSD12}) for $\lambda$ and
$\rho_s$ where we get
\begin{equation}\label{onelambda}
\lambda=1-C\cosh{2\theta^{\prime\prime}},
\end{equation}
\begin{equation}\label{onetheta}
\tanh{2(\theta^{\prime}-\theta^{\prime\prime})}
=\frac{(\lambda_1+\lambda_2)\sinh{2(\theta-\theta^{\prime\prime})}}
{(\lambda_1+\lambda_2)\cosh{2(\theta-\theta^{\prime\prime})} -C},
\end{equation}
\begin{equation}\label{onelambda12}
\lambda_{1,2}^{\prime}=\frac{1}{2\lambda}\left(
\frac{(\lambda_1+\lambda_2)\cosh{2(\theta-\theta^{\prime\prime})}
-C} {\cosh{2(\theta^\prime-\theta^{\prime\prime})}}
\pm(\lambda_3+\lambda_4)\right),
\end{equation}
\begin{equation}\label{onelambda34}
\lambda_j^{\prime}=\frac{\lambda_j}{\lambda}, \qquad
\mbox{for}\;\; j=3,4,
\end{equation}
where in Eq. (\ref{onelambda}) we have used
$\lambda_1^{\prime\prime}=\frac{1}{\cosh{2\theta^{\prime\prime}}}$
which follows from normalization condition of $\rho_e$. Finally
from the positivity conditions for $\lambda$ and $\lambda_i$ we
see that the following inequalities should hold
\begin{equation}
\cosh{2\theta^{\prime\prime}}\le \frac{1}{C}, \qquad
\cosh{2(\theta-\theta^{\prime\prime})}\le
\frac{\lambda_1-\lambda_2}{\lambda_1+\lambda_2}
+\frac{2\lambda_1\lambda_2}{(\lambda_1+\lambda_2)C}.
\end{equation}
Note that thus obtained decomposition is not a special case of the
decomposition considered in subsection \ref{subsecGwoot}. There we
considered the case that all $\rho$, $\rho_s$ and $\rho_e$ were
expressed in same Wootters basis. Here their Wootters basis
parameterized differently, namely $\theta$, $\theta^{\prime}$ and
$\theta^{\prime\prime}$ respectively. The optimal decomposition
given by Eqs. (\ref{onelambda}) to (\ref{onelambda34}) involves
some interesting special cases as follows

{\bf case i) $\theta=\theta^{\prime}$:} In this case from Eqs.
(\ref{onelambda}) to (\ref{onelambda34}) we get
$\theta^{\prime\prime}=\theta$, which yields the results of
subsection \ref{subsecGwoot} for a one parameter Wootters basis.

{\bf case ii) $\theta=0$, $\theta^{\prime\prime}\ne 0$:} This case
leads to  optimal decomposition of a BD state in terms of the non
maximal entangled pure state. This case also can be considered as
generalization of the result of Ref. \cite{shi}.

{\bf case iii) $\theta\ne 0$, $\theta^{\prime\prime}=0:$} This
case leads to  the optimal decomposition of a one parameter LOCC
transformed BD state in terms of maximal entangled pure state.

\subsection{Three parameters LOCC transformed BD states}
\label{subsecThreepara}  Now we consider the case that $\rho$ can
be obtained from BD states via three parameters LOCC
transformation as $\rho=Y\Lambda Y^\dag$ with \cite{jaf1}
\begin{equation}
{\small
Y=\left(\begin{array}{cccc}
\cosh{\theta}\cosh{\xi}\cosh{\phi} +\sinh{\theta}\sinh{\phi} &
i(\cosh{\theta}\cosh{\xi}\sinh{\phi} +\sinh{\theta}\cosh{\phi}) &
i\cosh{\theta}\sinh{\xi} & 0 \\
-i(\sinh{\theta}\cosh{\xi}\cosh{\phi} +\cosh{\theta}\sinh{\phi}) &
\sinh{\theta}\cosh{\xi}\sinh{\phi} +\cosh{\theta}\cosh{\phi} &
\sinh{\theta}\sinh{\xi} & 0 \\
-i\sinh{\xi}\cosh{\phi} & \sinh{\xi}\sinh{\phi} & \cosh{\xi} & 0
\\
0 & 0 & 0 & 1
\end{array}\right)},
\end{equation}
where normalization condition leads to
$$
Tr(\rho)=\left(\left(\lambda_1\cosh^2{\phi}+\lambda_2\sinh^2{\phi}\right)\cosh^2{\xi}
+\lambda_3\sinh^2{\xi}
+\left(\lambda_1\sinh^2{\phi}+\lambda_2\cosh^2{\phi}\right)\right)\cosh{2\theta}
$$
\begin{equation}
+\left(\lambda_1\cosh^2{\phi}+\lambda_2\sinh{\phi}\right)\sinh^2{\xi}
+\lambda_3\cosh^2{\xi}
+(\lambda_1+\lambda_2)\cosh{\xi}\sinh{2\theta}\sinh{2\phi}+\lambda_4=1.
\end{equation}
We choose below $\rho_s$ in the same form as $\rho$, i.e.
$\rho_s=Y^{\prime}\Lambda^{\prime}{Y^{\prime}}^\ast$ where
$\Lambda^{\prime}=
\mbox{diag}(\lambda_1^{\prime},\lambda_2^{\prime},
\lambda_3^{\prime},\lambda_4^{\prime})$ and $Y^{\prime}$ are
defined as $Y$ but here $\theta$, $\xi$ and $\phi$ are replaced
with $\theta^{\prime}$, $\xi^{\prime}$ and $\phi^{\prime}$. Now to
obtain optimal L-S decomposition we should take a generic density
matrix for $\rho_e$. It can be easily seen that Eq. (\ref{LSD})
requires that $\rho_e$ has also the same form as $\rho$ and
$\rho_s$. So we get
$\rho_e=Y^{\prime\prime}\Lambda^{\prime\prime}{Y^{\prime\prime}}^\ast$
where $\Lambda^{\prime\prime}=
\mbox{diag}(\lambda_1^{\prime\prime},\lambda_2^{\prime\prime},
\lambda_3^{\prime\prime},\lambda_4^{\prime\prime})$ and
$Y^{\prime\prime}$ is defined as $Y$ but here $\theta$, $\xi$ and
$\phi$ are replaced with $\theta^{\prime\prime}$,
$\xi^{\prime\prime}$ and $\phi^{\prime\prime}$. By using the above
considerations and Eq. (\ref{LSD}) we get
\begin{equation}\label{threeLSD1}
Y\Lambda Y^\dag=\lambda (Y^{\prime}\Lambda^{\prime}
{Y^{\prime}}^\dag)+(1-\lambda)
(Y^{\prime\prime}\Lambda^{\prime\prime} {Y^{\prime\prime}}^\dag).
\end{equation}
Now multiplying Eq. (\ref{threeLSD1}) by ${Y^{\prime\prime}}^T$
and ${Y^{\prime\prime}}^\ast$, respectively from left and right
and using the orthogonality of $Y^{\prime\prime}$ we get
\begin{equation}\label{threeLSD2}
({Y^{\prime\prime}}^T Y)\Lambda
(Y^\dag{Y^{\prime\prime}}^\ast)=\lambda ({Y^{\prime\prime}}^T
Y^{\prime})\Lambda^{\prime}
({Y^{\prime}}^\dag{Y^{\prime\prime}}^\ast)+(1-\lambda)\lambda^{\prime\prime},
\end{equation}
Subtracting three last diagonal elements of matrix equation
(\ref{threeLSD2}) from the first one and using the fact that
$\rho_s$ has zero concurrence, i.e.
$\lambda_1^{\prime}-\lambda_2^{\prime}-\lambda_3^{\prime}-\lambda_4^{\prime}=0$,
we get after some algebraic calculations
$\lambda=\frac{C^{\prime\prime}-C}{C^{\prime\prime}}$ and
$\frac{\rm{d}\lambda}{\rm{d}C^{\prime\prime}}=\frac{C}{{C^{\prime\prime}}^2}\ge
0$ where $C=\lambda_1-\lambda_2-\lambda_3-\lambda_4$ and
$C^{\prime\prime}=\lambda_1^{\prime\prime}-\lambda_2^{\prime\prime}
-\lambda_3^{\prime\prime}-\lambda_4^{\prime\prime}$, are
concurrence of $\rho$ and $\rho_e$, respectively. This shows that
maximal $\lambda$ is achieved when
$\lambda_2^{\prime\prime}=\lambda_3^{\prime\prime}=\lambda_4^{\prime\prime}=0$,
i.e. $\rho_e$ is pure entangled state with concurrence
$\lambda_1^{\prime\prime}$. Considering the above results we can
write Eq. (\ref{threeLSD1}) as
\begin{equation}
\rho_{11}=\lambda\rho_{11}^\prime+(1-\lambda)\lambda_1^{\prime\prime}
\left(\cosh{\theta}\cosh{\xi}\cosh{\phi}+\sinh{\theta}\sinh{\phi}\right)^2,
\end{equation}
\begin{equation}
\rho_{22}=\lambda\rho_{22}^\prime+(1-\lambda)\lambda_1^{\prime\prime}
\left(\sinh{\theta}\cosh{\xi}\cosh{\phi}+\cosh{\theta}\sinh{\phi}\right)^2,
\end{equation}
\begin{equation}
\rho_{33}=\lambda\rho_{33}^\prime+(1-\lambda)\lambda_1^{\prime\prime}
\sinh^2{\xi}\cosh^2{\phi},
\end{equation}
\begin{equation}
\rho_{44}=\lambda\rho_{44}^\prime,
\end{equation}
\begin{equation}
\rho_{12}=\lambda\rho_{12}^\prime
+(1-\lambda)\lambda_1^{\prime\prime}
\left(\left(\cosh^2{\xi}\cosh^2{\phi}
+\sinh^2{\phi}\right)\sinh{2\theta} +\cosh{\xi}\cosh{2\theta}
\sinh{2\phi}\right),
\end{equation}
\begin{equation}
\rho_{13}=\lambda\rho_{13}^\prime
+(1-\lambda)\lambda_1^{\prime\prime}
\left(\cosh{\theta}\cosh^2{\phi}\sinh{2\xi}
+\sinh{\theta}\sinh{\xi}\sinh{2\phi}\right),
\end{equation}
\begin{equation}
\rho_{23}=\lambda\rho_{23}^\prime
+(1-\lambda)\lambda_1^{\prime\prime}
\left(\sinh{\theta}\cosh^2{\phi}\sinh{2\xi}
+\cosh{\theta}\sinh{\xi}\sinh{2\phi}\right),
\end{equation}
where
$$
\rho_{11}=\left(\lambda_1\left( \cosh{\theta}\cosh{\xi}\cosh{\phi}
+\sinh{\theta}\sinh{\phi}\right)^2
+\lambda_2\left(\cosh{\theta}\cosh{\xi}\sinh{\phi}
+\sinh{\theta}\cosh{\phi}\right)^2\right.
$$
\begin{equation}
\left.+\lambda_3\left(\cosh{\theta}\sinh{\xi}\right)^2 \right),
\end{equation}
$$
\rho_{22}=\left(\lambda_1\left( \sinh{\theta}\cosh{\xi}\cosh{\phi}
+\cosh{\theta}\sinh{\phi}\right)^2
+\lambda_2\left(\sinh{\theta}\cosh{\xi}\sinh{\phi}
+\cosh{\theta}\cosh{\phi}\right)^2\right.
$$
\begin{equation}
\left.+\lambda_3\left(\sinh{\theta}\sinh{\xi}\right)^2 \right),
\end{equation}
\begin{equation}
\rho_{33}=\left(\lambda_1\sinh^2{\xi}\cosh^2{\phi}
+\lambda_2\sinh^2{\xi}\sinh^2{\phi}+\lambda_3\cosh^2{\xi}\right),
\end{equation}
\begin{equation}
\rho_{44}=\lambda_4,
\end{equation}
$$
\rho_{12}=\left(\left(\lambda_1\left(\cosh^2{\xi}\cosh^2{\phi}
+\sinh^2{\phi}\right) +\lambda_2\left(\cosh^2{\xi}\sinh^2{\phi}
+\cosh^2{\phi}\right)
+\lambda_3\sinh^2{\xi}\right)\sinh{2\theta}\right.
$$
\begin{equation}
+\left.(\lambda_1+\lambda_2)\cosh{\xi}\sinh{2\phi}
\cosh{2\theta}\right),
\end{equation}
\begin{equation}
\rho_{13}=\left(\left(\lambda_1\cosh^2{\phi}
+\lambda_2\sinh^2{\phi} +\lambda_3\right)\cosh{\theta}\sinh{2\xi}
+(\lambda_1+\lambda_2)\sinh{\theta}\sinh{\xi} \sinh{2\phi}\right),
\end{equation}
\begin{equation}
\rho_{23}=\left(\left(\lambda_1\cosh^2{\phi}
+\lambda_2\sinh^2{\phi} +\lambda_3\right)\sinh{\theta}\sinh{2\xi}
+(\lambda_1+\lambda_2)\cosh{\theta}\sinh{\xi}
\sinh{2\phi^{\prime}}\right),
\end{equation}
and $\rho_{ij}^\prime$ are defined in same form as $\rho_{ij}$ but
here all parameters are expressed in terms of prime parameters.
After tedious but straightforward calculations we arrive at the
following results for $\rho_s$
\begin{equation}\label{tanht}
\tanh{\xi^{\prime}}=\frac{-F\sinh{\theta^{\prime}}+G\cosh{\theta^{\prime}}}
{(p_1+p_2-A)\sinh{2\theta^{\prime}}+E\cosh{2\theta^{\prime}}},
\end{equation}
\begin{equation}\label{tanh2t}
\tanh{2\xi^{\prime}}=\frac{-F\cosh{\theta^{\prime}}+G\sinh{\theta^{\prime}}}
{p_1\cosh^2{\theta^{\prime}}+p_2\sinh^2{\theta^{\prime}}+p_3
-\frac{1}{2}(A\cosh{2\theta^{\prime}}-E\sinh{2\theta^{\prime}}+B+2D)},
\end{equation}
\begin{equation}
\tanh{2\phi^{\prime}}=\frac{F\sinh{\theta^{\prime}}-G\cosh{\theta^{\prime}}}
{\sinh{\xi^{\prime}}(\Lambda\lambda_3^{\prime}+(p_1+p_2-A)\cosh{2\theta^{\prime}}
-p_3+E\sinh{2\theta^{\prime}}+D)},
\end{equation}
\begin{equation}
\lambda_3^{\prime}=\frac{1}{2\lambda}\left(
\frac{-F\cosh{\theta^{\prime}}+G\sinh{\theta^{\prime}}}
{\sinh{2\xi^{\prime}}}-p_1\cosh^2{\theta^{\prime}}
-p_2\sinh^2{\theta^{\prime}}+P_3
+\frac{1}{2}(A\cosh{2\theta^{\prime}}-E\sinh{2\theta^{\prime}}+B-2D)\right),
\end{equation}
\begin{equation}
\lambda_1^{\prime}=\frac{1}{2\lambda}\left(
\frac{1}{\cosh{2\phi^{\prime}}}(\Lambda\lambda_3^{\prime}
+(p_1+p_2-A)\cosh{2\theta^{\prime}}-P_3
+E\sinh{2\theta^{\prime}}+D)+\Lambda\lambda_3^{\prime}+p_1-p_2-p_3-B+D\right),
\end{equation}
\begin{equation}
\lambda_2^{\prime}=\frac{1}{2\lambda}\left(
\frac{1}{\cosh{2\phi^{\prime}}}(\Lambda\lambda_3^{\prime}
+(p_1+p_2-A)\cosh{2\theta^{\prime}}-P_3
+E\sinh{2\theta^{\prime}}+D)-\Lambda\lambda_3^{\prime}-p_1+p_2+p_3+B-D\right),
\end{equation}
\begin{equation}
\lambda_4^{\prime}=\frac{p_4}{\lambda},
\end{equation}
where
\begin{equation}
A=(1-\lambda)\lambda_1^{\prime\prime}
\left((\cosh^2{\xi}\cosh^2{\phi}+\sinh^2{\phi})\cosh{2\theta}
+\cosh{\xi}\sinh{2\theta}\sinh{2\phi}\right),
\end{equation}
\begin{equation}
B=(1-\lambda)\lambda_1^{\prime\prime}
\left(\cosh^2{\xi}\cosh^2{\phi}-\sinh^2{\phi}\right),
\end{equation}
\begin{equation}
D=(1-\lambda)\lambda_1^{\prime\prime} \sinh^2{\xi}\cosh^2{\phi},
\end{equation}
\begin{equation}
E=(1-\lambda)\lambda_1^{\prime\prime}
\left((\cosh^2{\xi}\cosh^2{\phi}+\sinh^2{\phi})\sinh{2\theta}
+\cosh{\xi}\cosh{2\theta}\sinh{2\phi}\right)-\rho_{12},
\end{equation}
\begin{equation}
F=(1-\lambda)\lambda_1^{\prime\prime}
\left(\cosh{\theta}\cosh^2{\phi}\sinh{2\xi}
+\sinh{\theta}\sinh{\xi}\sinh{2\phi}\right)-\rho_{13},
\end{equation}
\begin{equation}
G=(1-\lambda)\lambda_1^{\prime\prime}
\left(\sinh{\theta}\cosh^2{\phi}\sinh{2\xi}
+\cosh{\theta}\sinh{\xi}\sinh{2\phi}\right)-\rho_{23}.
\end{equation}
The parameters $\theta^{\prime}$ and $\xi^{\prime}$ are obtained
by solving the  Eqs. (\ref{tanht}) and (\ref{tanh2t}), using the
remaining equations we can determine the parameters of $\rho_s$ in
terms of parameters of $\rho$ and $\rho_e$. Note that the one
parameter density matrix which was considered in previous
subsection can be obtain from three parameters one by setting
$\phi=\phi^{\prime}=\phi^{\prime\prime}=
\xi=\xi^{\prime}=\xi^{\prime\prime}=0$. One can see that the
equations in one parameter case is solvable and we can express the
parameters of separable and entangled parts in L-S decomposition
in terms of parameters of density matrix $\rho$ which is the
reason for its separated consideration in previous subsection.

\subsection{ $2\otimes 3$ Bell decomposable state}
\label{subsecBDS23}
In this subsection we obtain optimal L-S
decomposition for  Bell decomposable states of $2\otimes 3$
quantum systems. A Bell decomposable density matrix acting on
$2\otimes3$ Hilbert space can be defined by
\begin{equation} \label{BDS23}
\rho=\sum_{i=1}^{6}p_{i}\left|\psi_i\right>\left<\psi_i\right|,\quad\quad
0\leq p_i\leq 1,\quad \sum_{i=1}^{6}p_i=1,
\end{equation}
where $\left|\psi_i\right>$ are Bell states in $H^6\cong
H^2\otimes H^3$ Hilbert space, defined by:

$$
\left|\psi_1\right>=
\frac{1}{\sqrt{2}}(\left|11\right>+\left|22\right>), \qquad
\left|\psi_2\right>=
\frac{1}{\sqrt{2}}(\left|11\right>-\left|22\right>),
$$
\begin{equation}\label{BS123456}
\left|\psi_3\right>=
\frac{1}{\sqrt{2}}(\left|12\right>+\left|23\right>), \qquad
\left|\psi_4\right>=
\frac{1}{\sqrt{2}}(\left|12\right>-\left|23\right>),
\end{equation}
$$ \left|\psi_5\right>=
\frac{1}{\sqrt{2}}(\left|13\right>+\left|21\right>), \qquad
\left|\psi_6\right>=
\frac{1}{\sqrt{2}}(\left|13\right>-\left|21\right>). $$ It is
quite easy to see that the above states are orthogonal and hence
it can  span the Hilbert space of $2\otimes3$ systems. From
Peres-Horodeckis \cite{peres,horo0} criterion for separability we
deduce that the state given in Eq. (\ref{BDS23}) is separable if
and only if the following inequalities are satisfied
\begin{equation}\label{S1}
(p_1-p_2)^2\le(p_3+p_4)(p_5+p_6),
\end{equation}
\begin{equation}\label{S2}
(p_3-p_4)^2\le(p_5+p_6)(p_1+p_2),
\end{equation}
\begin{equation}\label{S3}
(p_5-p_6)^2\le(p_1+p_2)(p_3+p_4).
\end{equation}
In the sequel we always assume without loss of generality that
$p_1\ge p_2$, $p_3 \ge p_4$ and $p_5 \ge p_6$. Recently in Ref.
\cite{loz} an analytical lower bound of concurrence of any
$2\otimes K$ mixed state is derived as
\begin{equation}
C(\rho)\ge \sqrt{\sum_{i>j} C^2(\rho^{(ij)})},
\end{equation}
where $\rho^{(ij)}$ are unnormalized states restricted to
$2\otimes 2$ subsystems under projection operators $P^{(ij)}$ as
\begin{equation}
\rho^{(ij)}=P^{(ij)}\rho P^{(ij)}, \qquad
P^{(ij)}=I_2\otimes(\left|i\right>\left<i\right|
+\left|j\right>\left<j\right|),
\end{equation}
and $C(\rho^{(ij)})$ are Wootters concurrences of corresponding
restricted $2\otimes 2$ density matrices. For our $2\otimes 3$
Bell decomposable state we get
\begin{equation}\label{C12}
C(\rho^{(12)})=\rm{max}\{0,p_1-p_2-\sqrt{(p_3+p_4)(p_5+p_6)}\},
\end{equation}
\begin{equation}\label{C23}
C(\rho^{(23)})=\rm{max}\{p_3-p_4-\sqrt{(p_1+p_2)(p_5+p_6)}\},
\end{equation}
\begin{equation}\label{C13}
C(\rho^{(13)})=\rm{max}\{p_5-p_6-\sqrt{(p_1+p_2)(p_3+p_4)}\}.
\end{equation}
It is interesting to note that each Wootters concurrence given in
Eqs. (\ref{C12}) to (\ref{C13}) corresponds to separability
conditions given in Eqs. (\ref{S1}) to (\ref{S3}), respectively.
Now in order to obtain optimal L-S decomposition for BD state
given in Eq. (\ref{BDS23}) we choose $\rho_s=\sum_i
p_i^{\prime}\left|\psi_i\right>\left<\psi_i\right|$ and
$\rho_e=\sum_i
p_i^{\prime\prime}\left|\psi_i\right>\left<\psi_i\right|$. We also
assume without loss of generality that $\rho_s$ lies on the
separable-entangled boundary defined by (all other cases where
$\rho_s$ lies on other surfaces can be treated similarly)
\begin{equation}\label{SS1}
p_1^{\prime}-p_2^{\prime}
=\sqrt{(p_3^{\prime}+p_4^{\prime})(p_5^{\prime}+p_6^{\prime})}.
\end{equation}
Moreover $\rho_s$ must satisfies the other two separability
conditions (\ref{S2}) and (\ref{S3}). This means that entangled
state $\rho$ violates separability condition (\ref{S1}), i.e. we
have
\begin{equation}\label{E1}
p_1\ge p_2 +\sqrt{(p_3+p_4)(p_5+p_6)}.
\end{equation}
However, two other inequalities (\ref{S2}) and (\ref{S3}) may be
violated simultaneously. Taking into account the above
considerations and Eq. (\ref{LSD}) we get after some elementary
calculations the following equation
$$\hspace{-50mm}
(1-\lambda)^2\left((p_1^{\prime\prime}-p_2^{\prime\prime})^2
-(p_3^{\prime\prime}+p_4^{\prime\prime})
(p_5^{\prime\prime}+p_6^{\prime\prime})\right)
$$
$$
-(1-\lambda)\left(2(p_1-p_2)
(p_1^{\prime\prime}-p_2^{\prime\prime})
-(p_3+p_4)(p_5^{\prime\prime}+p_6^{\prime\prime})
-(p_5+p_6)(p_3^{\prime\prime}+p_4^{\prime\prime}) \right)
$$
\begin{equation}\label{EQ23}
\hspace{-56mm} +\left((p_1-p_2)^2-(p_3+p_4)(p_5+p_6)\right)=0.
\end{equation}
Below in the rest of this subsection we will use Eq. (\ref{EQ23})
to calculate $\lambda$ for some possibile values of
$p^{\prime\prime}_i, i=1,2,...,6$: as follows

{\bf i) $p^{\prime\prime}=1$:}

In this case Eq. (\ref{EQ23}) gives the following results
\begin{equation}\label{optLSD23-1}
\lambda=1-p_1-p_2+\sqrt{(p_3+p_4)(p_5+p_6)},\qquad
\rho_e=\left|\psi_1\right>\left<\psi_1\right|,
\end{equation}
\begin{equation}\label{optLSD23-2}
p_1^{\prime}=\frac{p_1-(1-\lambda)}{\lambda},\qquad
p_j^{\prime}=\frac{p_j}{\lambda}\quad
\mbox{for}\,\,j=2,\cdot\cdot\cdot,6.
\end{equation}
Furthermore $\rho_s$ must satisfies the separability conditions
(\ref{S2}) and (\ref{S3}) which leads to  the following
restrictions for $\rho$
\begin{equation}\label{LS23-1}
\begin{array}{l}
(p_3-p_4)^2\le
(p_5+p_6)\left(2p_2+\sqrt{(p_3+p_4)(p_5+p_6)}\right)
\\
(p_5-p_6)^2\le
(p_3+p_4)\left(2p_2+\sqrt{(p_3+p_4)(p_5+p_6)}\right)
\end{array}
\end{equation}
By using Eq. (\ref{E1}) one can see that conditions (\ref{LS23-1})
are stronger than separability conditions (\ref{S2}) and
(\ref{S3}), that is in this case only separability condition
(\ref{S1}) is violated by $\rho$. It is worth to mention that for
these states we are enable to give exact expression for
concurrence. As concurrence $C(\rho)$ is defined as the infimum
over all possible pure state decompositions, no decomposition can
have average concurrence smaller than $C(\rho)$. Since the
decomposition given by Eqs. (\ref{optLSD23-1}) and
(\ref{optLSD23-2}) constitute a maximal entangled pure state
$\left|\psi_1\right>$ and a separable state $\rho_s$, it follows
that its average concurrence is equal to the weight of entangled
part, namely $(1-\lambda)$. On the other hand for entangled states
restricted by equations (\ref{E1}) and (\ref{LS23-1}) we get
$C(\rho^{(12)})\ge 0$ and $C(\rho^{(13)})=C(\rho^{(13)})=0$. This
means that the lower bound is equal to $(1-\lambda)$, i.e.
\begin{equation}
C(\rho)=(1-\lambda)=p_1-p_2-\sqrt{(p_3+p_4)(p_5+p_6)}.
\end{equation}

{\bf ii) $p_1^{\prime\prime}+p_2^{\prime\prime}=1$:}

In this case by performing optimization procedure
$\frac{\partial\lambda}{\partial
p_1^{\prime\prime}}=\frac{\partial\lambda}{\partial
p_2^{\prime\prime}}=0$ in Eq. (\ref{EQ23}) (under constraint
$p_1^{\prime\prime}+p_2^{\prime\prime}=1$), we can see that thus
obtained equations from optimization procedure restrict the
density matrix $\rho$ to rank four one, namely $p_3=p_4=0$ or
$p_5=p_6=0$. Under this circumstances we get
$\lambda=\frac{C^{\prime\prime}-C}{C^{\prime\prime}}$ and
$\frac{\rm{d}\lambda}{\rm{d}C^{\prime\prime}}$, where $C$ and
$C^{\prime\prime}$ are concurrence of $\rho$ and $\rho_e$,
respectively. This means that maximum $\lambda$ happens when
$p_2^{\prime\prime}=0$ which  reduces to results of previous case.

{\bf iii) $p_1^{\prime\prime}+p_3^{\prime\prime}=1:$}

After optimization procedure with the constraint
$p_1^{\prime\prime}+p_3^{\prime\prime}=1$ we get
\begin{equation}
\begin{array}{c}
\lambda=1-(p_1-p_2)-(p_3+p_4)-\frac{1}{4}(p_5+p_6)
\\
p_1^{\prime}=\frac{2p_2-p_5-p_6}{2\lambda},\qquad
p_3^{\prime}=\frac{p_5+p_6-4p_4}{4\lambda},\qquad
p_j^{\prime}=\frac{p_j}{\lambda},\quad \mbox{for}\,\, j=2,4,5,6,
\end{array}
\end{equation}
where the following inequalities should be imposed in order
$\rho_s$ to be separable state
\begin{equation}\label{LS23-13}
\begin{array}{c}
2\left(p_4-\frac{1}{8}(p_5+p_6)\right)^2\le
(p_5+p_6)\left(p_2-\frac{1}{4}(p_5+p_6)\right),
\\
2(p_5-p_6)^2\le(p_5+p_6)\left(p_2-\frac{1}{4}(p_5+p_6)\right),
\\
4p_4\le p_5+p_6\le 2p_2.
\end{array}
\end{equation}

{\bf iv) $p_1^{\prime\prime}+p_5^{\prime\prime}=1$:}

Analogue to the case $p_1^{\prime\prime}+p_3^{\prime\prime}=1$ we
get
\begin{equation}
\begin{array}{c}
\lambda=1-(p_1-p_2)-\frac{1}{4}(p_3+p_4)-(p_5+p_6),
\\
p_1^{\prime}=\frac{2p_2-p_3-p_4}{2\lambda},\qquad
p_3^{\prime}=\frac{p_3+p_4-4p_6}{4\lambda},\qquad
p_j^{\prime}=\frac{p_j}{\lambda},\quad \mbox{for}\,\, j=2,3,4,6,
\end{array}
\end{equation}
whit restrictions
\begin{equation}\label{LS23-15}
\begin{array}{c}
2(p_3-p_4)^2\le(p_3+p_4)\left(p_2-\frac{1}{4}(p_3+p_4)\right),
\\
2\left(p_6-\frac{1}{8}(p_3+p_4)\right)^2\le
(p_3+p_4)\left(p_2-\frac{1}{4}(p_3+p_4)\right),
\\
4p_6\le p_3+p_4\le 2p_2.
\end{array}
\end{equation}

{\bf v)
$p_1^{\prime\prime}+p_3^{\prime\prime}+P_5^{\prime\prime}=1$:}

In this case it follows from optimization that rank $\rho$ should
be four, namely $p_4=p_6=0$. Under this conditions we get
\begin{equation}
\begin{array}{c}
\lambda=2p_2,  \\
p_1^{\prime}=p_2^{\prime}=\frac{1}{2},\qquad
p_3^{\prime}=p_4^{\prime}=p_5^{\prime}=p_6^{\prime}=0.
\end{array}
\end{equation}

\subsection{Werner states}\label{subsecWerner}
The Werner states are the only states that are invariant under
$U\otimes U$ operations. For $d\otimes d$ systems the Werner
states are defined by \cite{werner}
\begin{equation}
\rho_f=\frac{1}{d^3-d}\left((d-f)I+(df-1)F\right), \qquad -1\le
f\le 1,
\end{equation}
where $I$ stands for identity operator and $F=\sum_{i,j}\left|i
j\right>\left<j i \right|$. It is shown that Werner state is
separable iff $0\le f\le 1$. Now to obtain L-S decomposition for
Werner states we choose $\rho_{f=0}$ as separable part and
$\rho_{f^\prime}$ as entangled state , i.e. $\rho_f=\lambda
\rho_{f=0}+(1-\lambda)\rho_{f^\prime}$. Then from Eq. (\ref{LSD})
we get $\lambda=\frac{f^\prime-f}{f^\prime}$ and
$\frac{\rm{d}\lambda}{\rm{d}f^\prime}=\frac{f}{{f^\prime}^2}\le
0$, that is $\lambda$ is maximum when $f^\prime=-1$. Using the
above results we get
\begin{equation}
\lambda=f+1, \qquad \rho_e=\frac{1}{d(d-1)}\left(I-F\right).
\end{equation}

\subsection{Isotropic states}\label{subsecIsotropic}
The isotropic states are the only ones that are invariant under
$U\otimes U^\ast$ operations, where $^\ast$ denotes complex
conjugation. The isotropic states of $d\otimes d$ systems are
defined by \cite{horo3}
\begin{equation}
\rho_F=\frac{1-F}{d^2-1}\left(I-\left|\psi^{+}\right>\left<\psi^{+}\right|\right)
+F\left|\psi^{+}\right>\left<\psi^{+}\right| , \qquad 0\le F\le 1,
\end{equation}
where $\left|\psi^{+}\right>=\frac{1}{\sqrt{d}}\sum_{i}\left|i i
\right>$ is maximally entangled state. It is shown that isotropic
state is separable when $0\le F\le \frac{1}{d}$ \cite{horo3}. Now
in order to obtain optimal L-S decomposition we choose boundary
isotropic separable state with $F=1/d$ as separable part and
$\rho_{F^\prime}$ as entangled state where we get
$\lambda=\frac{d(F^\prime-F)}{dF^\prime-1}$ and
$\frac{\rm{d}\lambda}{\rm{d}F^\prime}=\frac{d^2(F-1/d)}{(dF^\prime-1)^2}\ge
0$, that is, $\lambda$ is maximum when $F^\prime=1$. Using the
above results we get
\begin{equation}\label{isoLSD}
\lambda=\frac{d(1-F)}{d-1}, \qquad
\rho_e=\left|\psi^{+}\right>\left<\psi^{+}\right|.
\end{equation}
It is interesting to stress that the average I-concurrence of the
decomposition (\ref{isoLSD}) is equal to the I-concurrence of the
state obtained in Ref. \cite{rungta2}. By using Eq. (\ref{Icon})
one can easily see that
$C(\left|\psi^{+}\right>)=\sqrt{2(1-1/d)}$, which can be used to
evaluate average I-concurrence of the decomposition
\begin{equation}\label{isoavecon}
(1-\lambda)C(\left|\psi^{+}\right>)=\sqrt{\frac{2d}{d-1}}\left(F-\frac{1}{d}\right),
\qquad \rm{for}\quad \frac{1}{d}\le F \le 1,
\end{equation}
which is equal to the I-concurrence of isotropic states which has
been obtained in \cite{rungta2}.

\subsection{  One parameter $3\otimes 3$ state}\label{subsec33}
Finally let us consider a one parameter state acting on $H^9\cong
H^3\otimes H^3$ Hilbert space as \cite{horo2}
\begin{equation}
\rho_{\alpha}=\frac{2}{7}\left|\psi^{+}\right>\left<\psi^{+}\right|+\frac{\alpha}{7}\sigma_{+}
+\frac{5-\alpha}{7}\sigma_{-}, \qquad 2\le \alpha \le 5,
\end{equation}
where
\begin{equation}
\begin{array}{l}
\left|\psi^{+}\right>=
\frac{1}{\sqrt{3}}\left(\left|11\right>+\left|22\right>+\left|33\right>\right),
\\
\sigma_{+}=\frac{1}{3}\left(\left|12\right>\left<12\right|
\left|23\right>\left<23\right|+\left|31\right>\left<31\right|\right),
\\
\sigma_{-}=\frac{1}{3}\left(\left|21\right>\left<21\right|
\left|32\right>\left<32\right|+\left|13\right>\left<13\right|\right).
\end{array}
\end{equation}
$\rho_{\alpha}$ is separable iff $2 \le \alpha \le 3$, it is bound
entangled iff $3 \le \alpha \le 4$
 and it is distillable entangled state iff $4
\le \alpha \le 5$ \cite{horo2}. To obtain L-S decomposition for
$\rho_\alpha$ we choose boundary separable state with $\alpha=3$
as $\rho_s$ and $\rho_e=\rho_{\alpha^\prime}$. After some
calculations we get
$\lambda=\frac{\alpha-\alpha^\prime}{3-\alpha^\prime}$ and
$\frac{\rm{d}\lambda}{\rm{d}\alpha^\prime}=
\frac{\alpha-3}{(3-\alpha^\prime)^2}\ge 0$. So the optimal L-S
decomposition is achieved by choosing $\alpha^\prime=5$ and we get
\begin{equation}
\lambda=\frac{5-\alpha}{2}, \qquad
\rho_e=\frac{2}{7}\left|\psi^{+}\right>\left<\psi^{+}\right|+\frac{5}{7}\sigma_{+}.
\end{equation}

\subsection{ Multi partite isotropic states}\label{subsecMultiiso}
In this subsection we obtain optimal L-S decomposition for a
n-partite d-levels system. Let us consider the following mixture
of completely random state $\rho_0=I/d^n$ and maximally entangled
state $\left|\psi^{+}\right>$
\begin{equation}\label{multirho}
\rho(s)=(1-s)\frac{I}{d^n}+s\left|\psi^{+}\right>\left<\psi^{+}\right|,\qquad
0\le s \le 1,
\end{equation}
where $I$ denotes identity operator in $d^n$-dimensional Hilbert
space and
$\left|\psi^{+}\right>=\frac{1}{\sqrt{d}}\sum_{i=1}^{d}\left|ii\cdot\cdot\cdot
i\right>$. The separability properties of the state
(\ref{multirho}) is considered in Ref. \cite{pitt2}. It is shown
that the above state is separable iff
$s=s_0=\left(1+d^{n-1}\right)^{-1}$.

Now to obtain optimal L-S decomposition we choose $\rho(s_0)$ as
separable part and $\rho(s^{\prime})$ as entangled part. By using
Eq. (\ref{LSD}) we get
$\lambda=\frac{s^{\prime}-s}{s^{\prime}-\left(1+d^{n-1}\right)^{-1}}$
and
$\frac{\rm{d}\lambda}{\rm{d}s^{\prime}}=\frac{s-\left(1+d^{n-1}\right)^{-1}}
{\left(s^{\prime}-\left(1+d^{n-1}\right)^{-1}\right)^2}$. This
means that the maximum $\lambda$ achieved when $s^{\prime}=1$, so
we get
\begin{equation}
\lambda=\frac{(1-s)(1+d^{n-1})}{d^{n-1}},\qquad
\rho_e=\left|\psi^{+}\right>\left<\psi^{+}\right|.
\end{equation}

\section{Conclusion}\label{secConclusion}
We have shown that for a given bipartite density matrix and by
choosing a suitable separable set on the separable-entangled
boundary, optimal Lewenstein-Sanpera decomposition can be obtained
via optimization over a generic entangled density matrix. Based on
this , optimal L-S decomposition is obtained for some bipartite
systems. We have obtained optimal decomposition for some bipartite
states such as $2\otimes 2$ and $2\otimes 3$ Bell decomposable
states, generic two qubit state in Wootters basis, iso-concurrence
decomposable states, states obtained from BD states via one
parameter and three parameters LOCC operations, $d\otimes d$
Werner and isotropic states, a one parameter $3\otimes 3$ state
and multi partite isotropic state. It is shown that in all
$2\otimes 2$ systems considered here  the average concurrence of
the decomposition is equal to the concurrence. We also obtain
exact expression  for concurrence of some $2\otimes 3$ BD states.
In the case of $d\otimes d$ isotropic states it is shown that the
average I-concurrence of the decomposition is equal to the
I-concurrence of the states. We conjecture that for all optimal
decomposition that entangled part is only a pure state, the
average I-concurrence of the decomposition is equal to the
I-concurrence of the state.

\end{document}